\documentclass[conference]{IEEEtran}
\IEEEoverridecommandlockouts
\usepackage{cite}
\usepackage{amsmath,amssymb,amsfonts}
\usepackage{algorithmic}
\usepackage{graphicx}
\usepackage{textcomp}

\usepackage{booktabs} % For formal tables

\usepackage{comment}
\usepackage{balance} 
\usepackage{booktabs} % For formal tables

\usepackage{blindtext}
\usepackage{wrapfig}
\usepackage{amsmath}
\usepackage[dvipsnames,svgnames,x11names]{xcolor}
\usepackage{soul}  
\definecolor{darkviolet}{rgb}{0.58, 0.0, 0.83}
\definecolor{chartreuse(web)}{rgb}{0.5, 1.0, 0.0}
\definecolor{darkpastelgreen}{rgb}{0.01, 0.75, 0.24}
\definecolor{bittersweet}{rgb}{1.0, 0.44, 0.37}
\newcommand{\notenes}[1]{\textcolor{darkviolet}{#1}}
\newcommand{\highlightenes}[1]{\textcolor{darkpastelgreen}{#1}}

\newcommand{\halfenes}{\includegraphics[width=.9em]{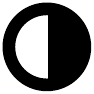}}
\newcommand{\fullenes}{\includegraphics[width=.9em]{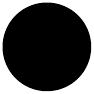}}
\newcommand{\emptyenes}{\includegraphics[width=.9em]{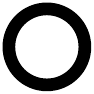}}

\newcommand{\yuvallak}[1]{\textcircled{{\scriptsize \textbf{#1}}}}

\usepackage{array}
\usepackage{ragged2e}
\newcolumntype{P}[1]{>{\RaggedRight\hspace{0pt}}p{#1}}
\newcolumntype{Z}[1]{>{\centering\arraybackslash\hspace{0pt}}p{#1}}

\newcommand*\rot{\rotatebox{90}}
\usepackage{pbox}

\begin{document}

\title{An Evaluation of Cryptocurrency Payment Channel Networks and Their Privacy Implications}%Privacy in off the Chain Payment Channel Networks}

\author{\IEEEauthorblockN{Enes Erdin\IEEEauthorrefmark{1},
Suat Mercan\IEEEauthorrefmark{2}, and Kemal Akkaya\IEEEauthorrefmark{2}} 
\IEEEauthorblockA{\IEEEauthorrefmark{1}Department of Computer Science, University of Central Arkansas, Conway, AR 72035, USA\\Email: \{eerdin\}@uca.edu}
\IEEEauthorblockA{\IEEEauthorrefmark{2}Dept. of Elec. and Comp. Engineering, Florida International University, Miami, FL 33174\\ Email: \{smercan,kakkaya\}@fiu.edu}
%\IEEEauthorblockA{\IEEEauthorrefmark{3}Dept. of Elec. and Comp. Engineering, Florida International University, Miami, FL 33174, Email: mncebe@gmail.com} 
%\IEEEauthorblockA{\IEEEauthorrefmark{4}Dept. of Elec. and Comp. Engineering, Florida International University, Miami, FL 33174, Email: kakkaya@fiu.edu} 
}

% \author{\IEEEauthorblockN{
% Enes Erdin,
% Suat Mercan and
% Kemal Akkaya}
% \IEEEauthorblockA{Dept. of Electrical and Computer Engineering \\Florida International University \\ Miami, Florida 33174\\
% Email: \{eerdi001, smercan, kakkaya\}@fiu.edu}}

% %% % % % % % % % 
% DEFINITIONS
\def\IdontNeedThis{1}

\maketitle
\begin{abstract}
 
Cryptocurrencies redefined how money can be stored and transferred among users. However, independent of the amount being sent, public blockchain-based cryptocurrencies suffer from high transaction waiting times and fees. These drawbacks hinder the wide use of cryptocurrencies by masses. To address these challenges, payment channel network  concept is touted as the most viable solution to be used for micro-payments. The idea is exchanging the ownership of money by keeping the state of the accounts locally. The users inform the blockchain rarely, which decreases the load on the blockchain. Specifically, payment channel networks can provide transaction approvals in seconds by charging a nominal fee proportional to the payment amount. Such attraction on payment channel networks inspired many recent studies which focus on how to design them and allocate channels such that the transactions will be secure and efficient. However, as payment channel networks are emerging and reaching large number of users, privacy issues are becoming more relevant that raise concerns about exposing not only individual habits but also businesses' revenues. In this paper, we first propose a categorization of the existing payment networks formed on top of blockchain-backed cryptocurrencies. After discussing several emerging attacks on user/business privacy in these payment channel networks, we qualitatively evaluate them based on a number of privacy metrics that relate to our case. Based on the discussions on the strengths and weaknesses of the approaches, we offer possible directions for research for the future of privacy based payment channel networks.

%Payment Channel networks are. Privacy in Payment Channel Networks are important. Public blockchain based cryptocurrencies do not scale well. They suffer from long confirmation times and high transaction fees during congestions. Among many solutions, off-the-chain payment mechanism is one of the promosing one and there have been studies in that area. We classified current stat-of-the art PCNs. We defined privacy attacks on user and business. We investigated current PCN wrt privacy they provide. We conclude the paper with possible research topics on privacy.

\end{abstract}

%     /$$$$$$                        /$$     /$$                    
%    /$$__  $$                      | $$    |__/                    
%   | $$  \__/  /$$$$$$   /$$$$$$$ /$$$$$$   /$$  /$$$$$$  /$$$$$$$ 
%   |  $$$$$$  /$$__  $$ /$$_____/|_  $$_/  | $$ /$$__  $$| $$__  $$
%    \____  $$| $$$$$$$$| $$        | $$    | $$| $$  \ $$| $$  \ $$
%    /$$  \ $$| $$_____/| $$        | $$ /$$| $$| $$  | $$| $$  | $$
%   |  $$$$$$/|  $$$$$$$|  $$$$$$$  |  $$$$/| $$|  $$$$$$/| $$  | $$
%    \______/  \_______/ \_______/   \___/  |__/ \______/ |__/  |__/
%
%\newpage%=============================================
%=====================================================
\section{Introduction}
\label{sec:introduction}

%In the evolving world of trade, the movement of money is going through changes. There are many modern money exchange systems. These can be classified into 4 major ones: the paper checks, credit/debit cards, automated clearing house (ACH) payments, and bank transfers.
There are many modern money exchange systems such as paper checks, credit/debit cards, automated clearing house (ACH) payments, bank transfers, or digital cash which are owned and regulated by financial institutions. Nevertheless, in the evolving world of trade, the movement of money is still going through changes. The last decade witnessed the introduction of \textit{Bitcoin} \cite{nakamoto2008bitcoin}, a new paradigm-shifting innovation where the users control their own money without needing a trusted third party. In this model, the users are governing the system by coming to a consensus for controlling the transfer and the ownership of the money. Following the success of Bitcoin, new cryptocurrencies that offer new capabilities were introduced based on the idea of consensus-based account management \cite{mukhopadhyay2016brief,tschorsch2016bitcoin}.

In the following years, the initial success of cryptocurrencies was hindered due to practical issues related its daily use. Basically, it was a very limited system in terms of scalability and its wide usage in simple daily transactions was quite impossible due to long waiting times, high disproportional transaction fees, and low throughput.
%These problem has been tackled by several studies over the last years.

% \begin{figure}
%     \centering
%     \includegraphics[width=.95\linewidth]{figs/AllLayers2.pdf}
%     \caption{Two users agree on opening a shared channel. They exchange money in this account and when the channel closes they get their parts.}
%     \label{fig:AllLayers}
% \end{figure}

Among many solutions \textit{payment channel} idea arose as a well-accepted one for solving mentioned problems. The idea is based on establishing off-chain links between parties so that many of the transactions would not be written to the blockchain each time. The payment channel idea later evolved towards the establishment of \textit{payment channel networks} (PCN), where among many participants and channels the participants pay by using others as relays, essentially forming a connected network. This is in essence a \textit{Layer-2} network application running on top of a cryptocurrency, which covers \textit{Layer-1} services. A perfect example to PCNs is Lightning Network (LN) \cite{poon2016bitcoin} which uses Bitcoin and reached to many users in a very short amount of time. Raiden \cite{Raiden}, based on Ethereum, is another example for a successful PCN.

The emergence of PCNs led to several research challenges. In particular, security of the off-chain payments is very important as users can lose money or liability can be denied. In addition, efficiency of payment routing within the PCN with large number of users is tackled. Such efforts paved the way for introducing many new PCNs in addition to LN. These PCNs rely on various cryptocurrencies and carry several new features. As these newly proposed PCNs become more prominent there will be a heavy user and business involvement which will raise issues regarding their privacy just as the user privacy on Internet. The difference is that on many cases, Internet privacy could be regulated but this will not be the case for PCNs as their very idea is based on decentralization. For instance, a user will naturally want to stay anonymous to the rest of the network while a business would like to keep its revenue private against its competitors.

%In this paper, we first give an introductory background on blockchain data structure and smart contracts. 

Therefore, in this paper, we investigate this very emerging issue and provide an analysis of current PCNs along with their privacy implications. We first categorize the PCNs in the light of common network architectures and blockchain types. %This is the first comprehensive categorization of PCNs. 
We then define user and business privacy within the context of PCNs, and discuss possible attacks on the privacy of the participants. % in the PCNs. 
Specifically, we came up with novel privacy risks specific to PCNs. Utilizing these attack scenarios, we later survey and evaluate thoroughly the existing PCNs in terms of their privacy capabilities based on certain metrics. This is a novel qualitative evaluation to be able to compare what each PCN is offering in terms of its privacy features.  Finally, we offer potential future research issues that can be further investigated in the context of PCN privacy. Our work not only is the first to increase awareness regarding the privacy issues in the emerging realm of PCNs but also will help practitioners on selecting the best PCN for their needs.   %extend our discussion as a contribution to the classification of state-of-the-art solutions for what they offer in terms of privacy. We conclude the paper with possible research directions.

%The paper is organized as follows: (KA: PLEASE PUT A BRIEF OUTLINE HERE). 

The paper is organized as follows: Section \ref{sec:background} gives an introductory background.
Next, Section \ref{sec:PCNCategorization} categorizes the PCNs in the light of common network architectures and blockchain types. In Section \ref{sec:PrivacyMetrics} we define the user and business privacy, discuss possible attacks on the privacy of the participants in the PCNs, and present an evaluation of state-of-the-art solutions for what they offer in terms of privacy. Section \ref{sec:FutureResearch} offers directions about the future research on privacy in PCNs and Section \ref{sec:conclusion} concludes the paper.

%We define the privacy from the user and the business perspective. 

%We define the attack types for the privacy concerns.

%We classify current state-of-the-art PCN networks.

%We survey current state-of-the art networks from the privacy point of view.

%We conclude paper for the possible research directions in the privacy of the PCNs.

%\begin{figure}
% \centering
% \includegraphics[width=.85\linewidth]{figs/PaymentScheme.pdf}
% \caption{DANGLING}
% \label{fig:my_label}
%\end{figure}

%     /$$$$$$                        /$$     /$$                    
%    /$$__  $$                      | $$    |__/                    
%   | $$  \__/  /$$$$$$   /$$$$$$$ /$$$$$$   /$$  /$$$$$$  /$$$$$$$ 
%   |  $$$$$$  /$$__  $$ /$$_____/|_  $$_/  | $$ /$$__  $$| $$__  $$
%    \____  $$| $$$$$$$$| $$        | $$    | $$| $$  \ $$| $$  \ $$
%    /$$  \ $$| $$_____/| $$        | $$ /$$| $$| $$  | $$| $$  | $$
%   |  $$$$$$/|  $$$$$$$|  $$$$$$$  |  $$$$/| $$|  $$$$$$/| $$  | $$
%    \______/  \_______/ \_______/   \___/  |__/ \______/ |__/  |__/
%
%\newpage%=============================================
%=====================================================
\section{Background}
\label{sec:background}

%\subsection{Blockchain and Smart Contracts}

\subsection{Blockchain}
\label{sec:Blockchain}
Blockchain is the underlying technology in cryptocurrency, that brings a new distributed database which is a public, transparent, persistent, and append-only ledger co-hosted by the participants. With various cryptographically verifiable methods, called \textit{Proof-of-X} (PoX), each participant in the network holds the power of moderation of the blockchain \cite{bano2019sok}. As an example, %being the first invented and largest cryptocurrency, 
Bitcoin and %the second largest one, 
Ethereum, which jointly hold 75\% of total market capitalization in the cryptocurrency world, utilize \textit{proof-of-work} (PoW) mechanism where a participant has to find a ``block-hash-value'' smaller than a jointly agreed number. A block is an element with a limited size that stores the transaction information. Each block holds the hash of the preceding block which in the long run forms a chain of blocks, called, the blockchain. ``Who-owns-what'' information is embedded in the blockchain as transaction information. Therefore, the cohort of independent participants turns blockchain into a liberated data/asset management technology free of trusted third parties. 

\subsection{Cryptocurrency}
Although it finds many areas, the most commonly used application of blockchain technology is cryptocurrencies. %, which also will interchangeably call money in this paper. 
A \textit{cryptocurrency} is a cryptographically secure and verifiable currency that can be used to purchase goods and services. In this paper, we will use cryptocurrency and money interchangeably.  

Blockchain technology undoubtedly changed the way data can be transferred, stored, and represented. Nonetheless, making a consensus on the final state of a distributed ledger has drawbacks. The first drawback is long transaction confirmation times. For example, in Bitcoin, a block is generated at about every 10 minutes. As a heuristic Bitcoin users wait 6 blocks for the finality of a transaction which yields almost 60 minutes.
%waiting time for finalizing a transaction.
In Ethereum, time between blocks are shorter but users wait 30 consecutive blocks which yields 10-15 minutes of waiting time. Note that, as a block is limited in size, not only the throughput will be limited, but also the total waiting time for the users will be longer during the congested times of the transfer requests. Nevertheless, if a user is in a hurry for approval of its transaction, it will need to pay larger fees to miners than what its competitors pay. This brings us the second drawback of using blockchain for cryptocurrency. The miner nodes, which generate and approve blocks, get fees from the users to include transactions in blocks.
%The fee amount is independent of the amount being transacted.
%During the high network congestion times, to make a larger profit, miners will be extremely selective in picking the requests from the transaction request pool (\textit{mempool}).
So when there is congestion, a payer either has to offer more fee or she/he has to wait more so that a miner picks her/his transaction request.% As indirectly explained, this is the reason for the longer confirmation times for regular users during network congestion. %(SM: PoW ITSELF IS ALSO ONE OF THE REASONS FOR LOW TRANSACTION RATE, I DON'T SEE WE MENTIONED THIS IF WE NEED TO)

\subsection{Smart Contracts}
%\cite{zhang2017cold} Social implications of networks like Zelle and so. They downloadede venmo public transactions.
The ability to employ smart contracts is another feature that makes blockchain an unorthodox asset management technology. Smart contracts are scripts or bytecodes, which define how transactions will take place based on the future events defined within the contract. Smart contracts can be utilized in conditional/unconditional peer-to-peer (P2P) transactions, voting, legal testament etc. As always, the duty of decision-making is on blockchain. Hence, the blockchain finalizes the transaction outputs when the smart contracts are utilized too. %While preparing a contract, the parties should be careful such that when the contract is awarded the transactions will not be reversible.

%     /$$$$$$                        /$$     /$$                    
%    /$$__  $$                      | $$    |__/                    
%   | $$  \__/  /$$$$$$   /$$$$$$$ /$$$$$$   /$$  /$$$$$$  /$$$$$$$ 
%   |  $$$$$$  /$$__  $$ /$$_____/|_  $$_/  | $$ /$$__  $$| $$__  $$
%    \____  $$| $$$$$$$$| $$        | $$    | $$| $$  \ $$| $$  \ $$
%    /$$  \ $$| $$_____/| $$        | $$ /$$| $$| $$  | $$| $$  | $$
%   |  $$$$$$/|  $$$$$$$|  $$$$$$$  |  $$$$/| $$|  $$$$$$/| $$  | $$
%    \______/  \_______/ \_______/   \___/  |__/ \______/ |__/  |__/
%
%===================================================
%\newpage%===========================================
%===================================================
\section{PCNs and their Categorization}
\label{sec:PCNCategorization}
%OR WE CAN SAY CLASSIFICATION BASED ON NETWORK AN BLOCKCHAIN ARCHITECTURES
%In this section, we provide details on PCNs and introduce a categorization to classify them based on some characteristics. 

\subsection{Payment Channel Networks}

%\notenes{Instead of pcn we can start with the payment channels or state channels, this is by the way only calssification done in the moreno sanchez paper.}

Due to the scalability issues researchers have always been in the search of solutions to make the cryptocurrency scalable. Among many offered solutions, the \textit{off-chain} payment channel idea has attracted the most interest. To establish such a channel, two parties agree on depositing some money in a multi-signature (2-of-2 multi-sig) wallet with the designated ownership of their share. The multi-sig wallet is created by a smart contract where both parties sign. The smart contract, mediated by the blockchain, includes the participants' addresses, their share in the wallet, and information on how the contract will be honored.
%The channel opening transaction is submitted to the blockchain.
Approval of the opening transaction in the blockchain initiates the channel. The idea is simple; the payer side gives ownership of some of his/her money to the other side by locally updating the contract mutually. To close the channel the parties submit ``closing transaction'' to the blockchain for it to honor the final state of the channel. Thus, each side receives its own share from the multi-sig wallet.% (SM: DO WE NEED TO MENTION HOW CHEATING IS PREVENTED?)

%When more of these channels come together, by each node having channels with other nodes, a payer can make transactions in a multi-hop fashion.   

\begin{figure}[htb]
   \centering
   \includegraphics[width=.85\linewidth]{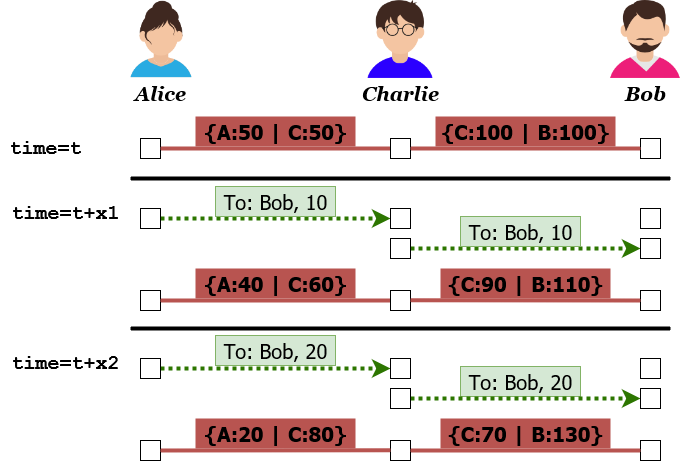}
   \caption{A Simple Multi-hop Payment. Alice can initiate a transfer to Bob utilizing channels between Alice-Charlie and Charlie-Bob.}
   \label{fig:PCN}
\end{figure}

Payment channels created among many parties make establishment of \textit{multi-hop payments} from a source to a destination through intermediary nodes possible.
%also have channels among each other.
As shown in Fig. \ref{fig:PCN}, Alice-Charlie (A-C) and Charlie-Bob (C-B) have channels. Let, A-C and C-B are initialized when \texttt{time} is \texttt{t}. Although Alice does not have a direct channel to Bob, she can still pay Bob via Charlie. At \texttt{time} \texttt{t+x1}, Alice initiates a transfer of 10 units to Bob. The money is destined to Bob over Charlie.
%\highlightenes{Along with the transaction request Alice gives the hash of a pre-image to Charlie. In order for Charlie to get the money from Alice he has to know the pre-image of that hash as a proof that Bob received the transaction.}
When Charlie honors this transaction in the C-B channel by giving 10 units to Bob, Alice gives 10 units of her share to Charlie in A-C channel. When the transfers are over, A-C and C-B channels get updated. When \texttt{time} is \texttt{t+x2}, Alice makes another transaction (20 units) to Bob and the shares in the channels get updated once again. 

\begin{figure}[htb]
   \centering
   \includegraphics[width=.85\linewidth]{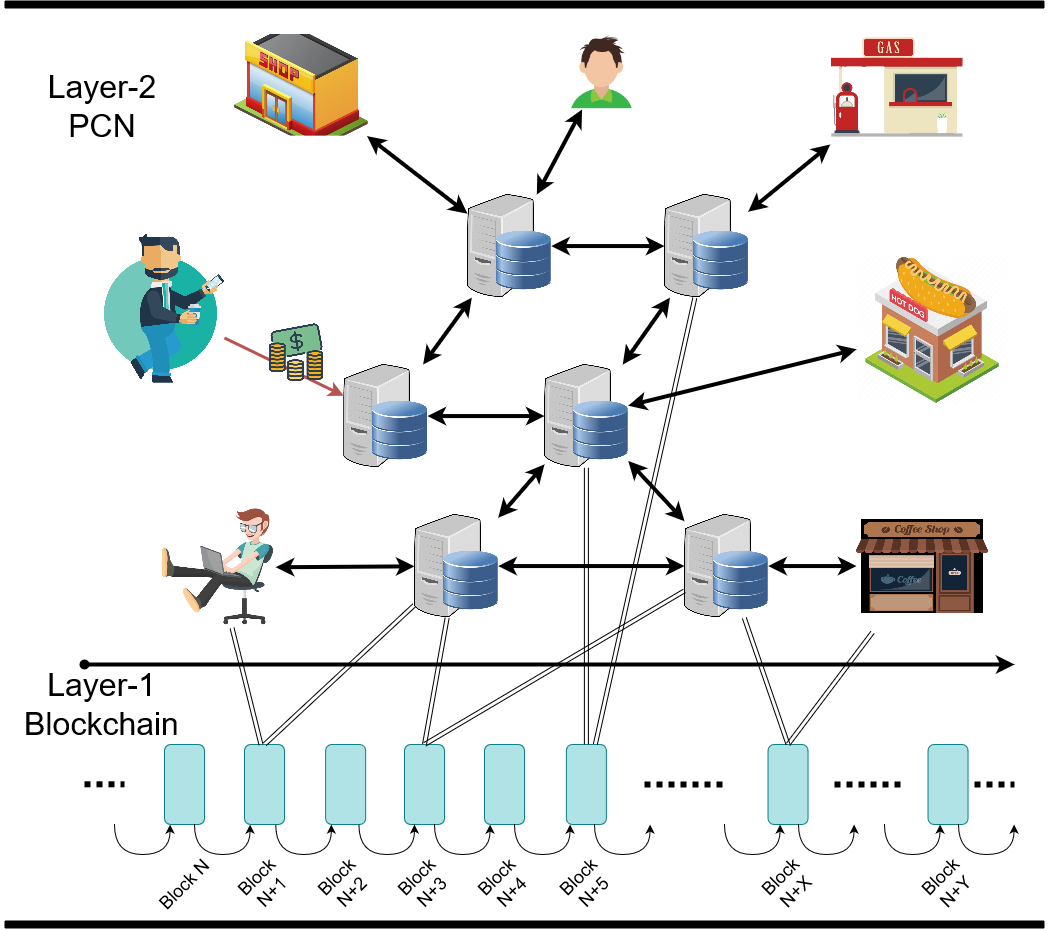}
   \caption{ %The users and businesses independently come together and establish payment channels between each other. Consequently, they form 
   A PCN of end-users and relays acting as backbone.} %Solid-arrowed lines represent channels between the nodes. Double lines represent how they agreed in the blockchain to establish a channel (only some of these are shown for simplicity).}% Some of these nodes might serve only for relaying (servers) while some of them might be either retail stores or humans sending and receiving. Here we see Bob is paying someone. (KA: THIS TEXT IS TOO LONG FOR AN INTRODUCTION FIGURE)}
   \label{fig:PCN-model}
\end{figure}

Multi-hop payment concept enables the establishment of a network of payment channels among users, which is referred to as PCN as shown in Fig. \ref{fig:PCN-model}. Current PCNs vary in terms of what topologies they depend on and which layer-1 blockchain technology they utilize. We discuss this categorization next. We will then explain each of these PCNs in more detail and categorize them in Section IV.  % (SM: I THINK WE CAN TALK ABOUT TRANSACTION FEE WHICH INCENTIVES OTHER NODES TO TRANSFER MONEY. NOT DIRECTLY RELATED TO THIS PAPER. HOW DO NODES ARE ENFORCED NOT TO CHARGE MORE FEE THAN THEY ANNOUNCED? )

\subsection{PCN Architectures}
In this section, we categorize the types of \textit{network architectures} that can be used in PCNs.

   \subsubsection{Centralized Architecture}
   In this type of network, there is a central node, and users communicate with each other either over that central node or based on the rules received from the central node as shown in Fig. \ref{fig:networkTypes}(a). From the governing point of view, if an organization or a company can solely decide on the connections, capacity changes, and flows in the network, then this architecture is called to be a centralized one.
 
   \subsubsection{Distributed Architecture}
   In distributed networks, there is no central node. As opposed to the centralized network, each user has the same connectivity, right to connect, and voice in the network. %, namely, the nodes and right to connect are distributed equally. 
   A sample architecture is shown in Fig. \ref{fig:networkTypes}(b).
 
   \subsubsection{Decentralized Architecture}
\begin{figure}[htb]
    \centering
    \includegraphics[width=.85\linewidth]{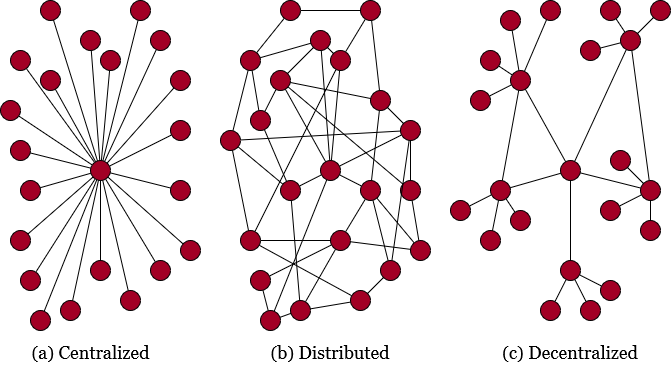}
    \caption{Network Types}
    \label{fig:networkTypes}
\end{figure}

   This type of architecture is a combination of the previous two types which is shown in Fig. \ref{fig:networkTypes}(c). In this architecture, there is no singular central node, but there are independent central nodes. When the child nodes are removed, central nodes' connections look very much like a distributed architecture. However, when the view is concentrated around one of the central nodes, a centralized architecture is observed.
   %it is seen that on a small scale there is a centralized architecture. %Most of the current PCNs can be classified under this type. Although the users are free to connect to any other node, we see that (e.g., Lightning Network) some nodes are popular (highly influential) and they have more channels than the rest.
 
   \subsubsection{Federated Architecture}
   Federated architecture sounds very much like the federation of the states in the real world and arguably lies somewhere between centralized and decentralized networks. In a federated architecture, there are many central nodes where they are connected to each other in a P2P fashion. Then the remaining nodes (children) strictly communicate with each other over these central nodes which very much looks like a federation of centralized architectures. %Moreover, each federation can come up with their local rules in addition to the protocol being used.% (SM: I THINK THIS IS KIND OF A LAYERED APPROACH. SIMILAR TO INTRA-AS, INTER-AS ROUTING)

\subsection{Types of Blockchain Networks}
In this section, we categorize the existing PCNs based on blockchain type they employ. %Note that this will help better understand the privacy approaches of the studies, their strength, and their weakness. 
There are mainly three types of blockchains employed by PCNs:

  \subsubsection{Public Blockchain}
  In a public blockchain, %as the name suggests, the blockchain where 
  no binding contract or registration is needed to be a part of the network. Users can join or leave the network whenever they want. Consequently, the PCN will be open to anyone who would like to use it. %All of the cryptocurrencies fall in this category of Blockchain.
 
  \subsubsection{Permissioned Blockchain}
  Permissioned (i.e., Private) blockchain lays at the opposite side of the public blockchain, where the ledger is managed by a company/organization. Moreover, the roles of the nodes within the network are assigned by the central authority. Not everybody can participate or reach to the resources in the permissioned blockchain. PCNs employing permissioned blockchain will be ``members-only''. %Blockchains used for supply chain management are good examples of this type.

  \subsubsection{Consortium Blockchain}
  Contrary to the permissioned blockchain, in consortium blockchain, the blockchain is governed by more than one organization. From the centralization point of view, this approach seems more liberal but the governance model of the blockchain slides it to the permissioned side. PCNs utilizing consortium blockchain will be similar to permissioned blockchain in terms of membership but in this case members will be approved by the consortium.
  %Facebook's draft cryptocurrency, Libra, is a good example of this type of blockchain.

%%%%%%%%%%%%%%%%%%%%%%%%%%%%%%%%%%%%%%%%
%%%%%%%%%%%%%%%%%%%%%%%%%%%%%%%%%%%%%%%%
%$%$%$%$%$%$%$%$%$%$%$%$%&%&%&%&%&%&%&%&

%     /$$$$$$                        /$$     /$$                    
%    /$$__  $$                      | $$    |__/                    
%   | $$  \__/  /$$$$$$   /$$$$$$$ /$$$$$$   /$$  /$$$$$$  /$$$$$$$ 
%   |  $$$$$$  /$$__  $$ /$$_____/|_  $$_/  | $$ /$$__  $$| $$__  $$
%    \____  $$| $$$$$$$$| $$        | $$    | $$| $$  \ $$| $$  \ $$
%    /$$  \ $$| $$_____/| $$        | $$ /$$| $$| $$  | $$| $$  | $$
%   |  $$$$$$/|  $$$$$$$|  $$$$$$$  |  $$$$/| $$|  $$$$$$/| $$  | $$
%    \______/  \_______/ \_______/   \___/  |__/ \______/ |__/  |__/
%
%\newpage%=============================================
%=====================================================
\section{Privacy Issues in PCNs: Metrics and Evaluation }
\label{sec:PrivacyMetrics}

As PCNs started to emerge within the last few years, a lot of research has been devoted to make them efficient, robust, scalable and secure. However, as some of these PCNs started to be deployed, they reached large number of users (i.e., LN has over 10K users), which is expected to grow further. % as long as users are satisfied with their services. 
Such growth brings several privacy issues that are specific to PCNs. %In this respect, we observed that strengthening the security in PCNs comes with weaker privacy while strengthening the privacy in PCN makes the network less practical.
We argue that %very little attention has been paid to these issues and 
there is a need to identify and understand privacy risks in PCNs from both the users and businesses perspectives.  Therefore, in this section, we first define these privacy metrics and explain possible privacy attacks in PCNs. We then summarize the existing PCNs to evaluate their privacy capabilities with respect to these metrics for the first time. %Our goal is to increase awareness to not only strengthen the privacy features of the existing PCNs but also help designers to consider privacy-by-design principle when creating new PCNs from scratch. % Next, we summarize the state-of-the-art PCN proposals.

%              _                   _   _             
%             | |                 | | (_)            
%    ___ _   _| |__  ___  ___  ___| |_ _  ___  _ __  
%   / __| | | | '_ \/ __|/ _ \/ __| __| |/ _ \| '_ \ 
%   \__ \ |_| | |_) \__ \  __/ (__| |_| | (_) | | | |
%   |___/\__,_|_.__/|___/\___|\___|\__|_|\___/|_| |_|
\subsection{Privacy in PCNs}

%In Bitcoin, privacy plays a vital role. Although, attacks related to de-anonymization in Bitcoin is out of scope in this paper, by utilizing pseudonyms, the real identities of the users were aimed to be kept private. It is seen that inherited from this philosophy, PCN designers also pay attention to privacy features with different points of view. Nevertheless, we observe that strengthening the security in PCN comes with weaker privacy or strengthening privacy makes the network less practical. In this paper, we seek answers for the following question; if PCNs were widely used for daily transactions, from web purchases to retailing, what kind of privacy would they provide, and to what extent?

In its simplest form, \textit{data privacy} or \textit{information privacy} can be defined as the process which answers how storage, access, and disclosure of data take place. %For centrally managed systems the central node (or company) is the responsible party for preserving the privacy of the users by defining appropriate policies to manage their data. However, when the system shifts towards a decentralized/distributed one, the privacy of the users should be taken care of by the protocol running beneath the network. 
The PCN, in our case, needs to provide services ensuring  that  the users' data will not be exposed without their authorization. However, the user data travels within the PCN through many other users. 
%Hence, in order to assess the level of privacy in a particular decentralized system, definitions for privacy within the system are needed.
%\notenes{https://fas.org/sgp/crs/ \\ secrecy/R43714.pdf}
%For instance, let us consider a payment in a PCN  to take place as shown in Fig. \ref{fig:TypicalNetwork} where the sender, $u_s$, initiates a transaction destined to the recipient, $u_r$. The payment passes through other nodes $u_{1...N}$ where $N\geq1$.
%Note that an intermediary node either can be defined as a normal node like others or it may be equipped with extra capabilities like directing payments or gathering information from other nodes.
%\highlightenes{(SM: SUCH AS? like landmarks explain this better)}.
%A node may have established extra channels (shown with blue dashed arrows) other than the channel it uses for a particular payment. Unless otherwise stated a payment channel is assumed to be bidirectional. As seen, the potential of exposure of the data to numerous intermediaries in PCNs is huge.  
%\begin{figure}[htb]
%   \centering
%   \includegraphics[width=.6\linewidth]{figs/PaymentScheme-TypicalNetwork.pdf}
%   \caption{In a typical example the sender $u_s$ initiates a transaction to $u_r$ through the intermediary nodes $u_{1...N}$. A node can open many channels.}
%   \label{fig:TypicalNetwork}
%\end{figure}
%In the state-of-the-art studies, the privacy notion, the attacker models, and their capabilities vary. 
To address these issues, some PCN works aimed to hide the sender ($u_s$) or receiver ($u_r$) identity (i.e., \textit{anonymity}) whereas some others concentrated on strengthening the \textit{relationship anonymity} between sender and recipient.
%among the payments and a sender or receiver.
%In the next section, we provide a summary of our assumptions and privacy attacks we considered in this paper. %Throughout this paper, we will consider different kinds of attacks based on the privacy metrics. Nonetheless, privacy totally depends on the needs of the users in the network. 

\subsection{Attack Model and Assumptions}

\begin{figure}[htb]
   \centering
   \includegraphics[width=.6\linewidth]{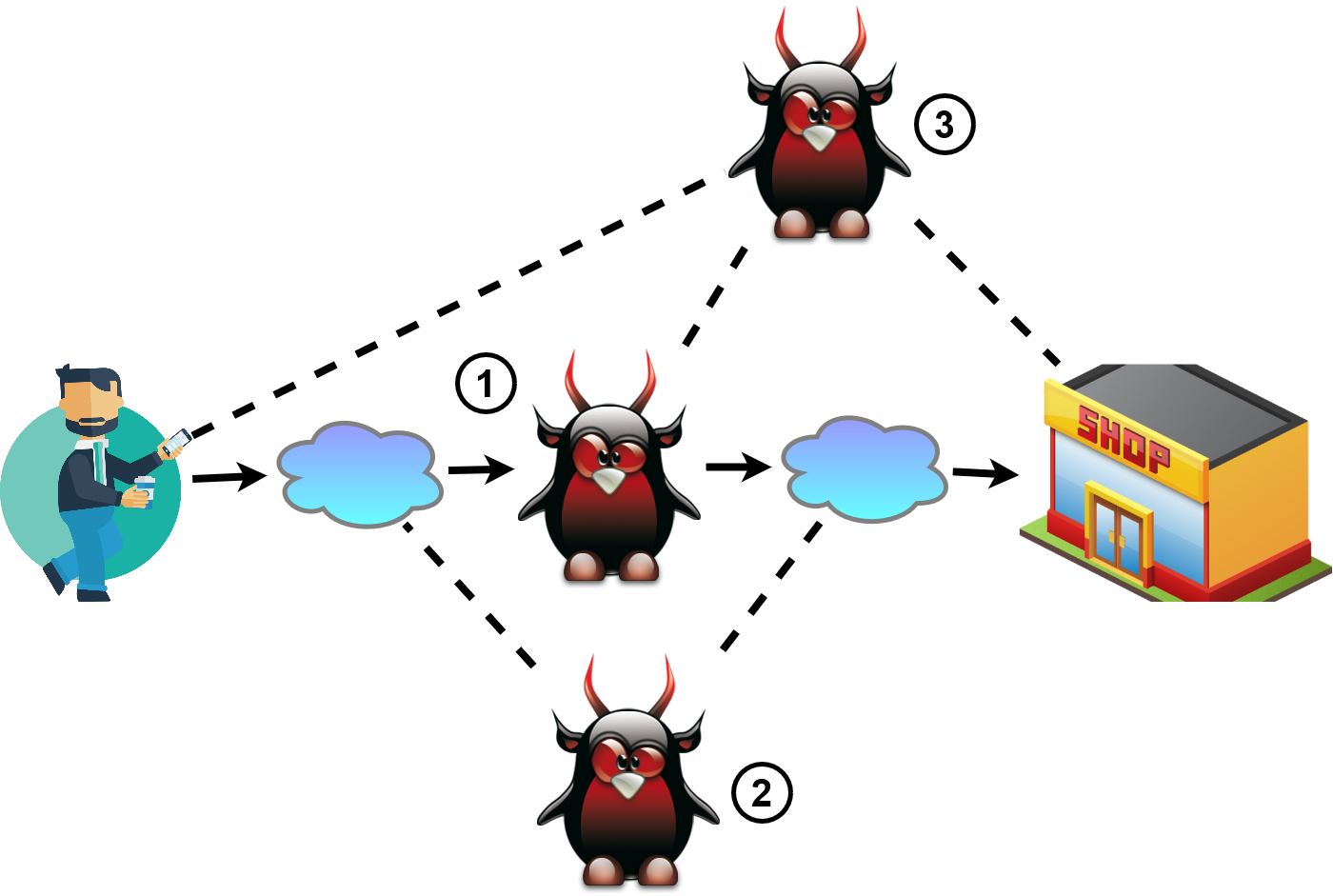}
   \caption{Attackers can appear in the network in different places. %\yuvallak{1} The attacker is on the path of a payment. \yuvallak{2} The attacker is not on the path of a particular payment but she can partially observe the changes in the network. \yuvallak{3} The attacker colludes with the other nodes.
   }
   \label{fig:AttackersConnected}
\end{figure}

There are two types of attackers considered in this paper. The first attacker is an \textit{honest-but-curious} (HBC) where the attacker acts honestly while running the protocols but still collects information passively during operations. The second attacker of interest is the \textit{malicious} attacker that controls more than one node in the network to deviate from the protocols. %Hence, it can act based on its own rules.
%, e.g. denial of service or colluding with other nodes in order to learn about the user/payment information.
%For both of the attacker types the attacker either tries to learn the origin and the destination of payment or tries to learn the path of the payment routing. This information can be used for a couple of purposes. The first purpose of trying to get this information is censoring the payment by simply rejecting it. The second purpose is trying to guess the business capacity of a node. The third reason is trying to learn the spending habits of customers. If a single item is purchased, a persistent attacker will be able to relate the payment to the service or good that has been purchased. The fourth purpose is trying to discredit a particular node simply by slowing down the transaction so that the customers will tend to lose interest in that seller because of a lack of payment usability.
%Attacks on privacy will eventually result in attacks on the security (e.g., denial of service attacks, censorship, hijacking) of the network but security attacks are out of scope within the context of this study. 
These attacker types and how they can situate in the network are shown in Fig. \ref{fig:AttackersConnected} as follows: \yuvallak{1} The attacker is on the path of a payment. \yuvallak{2} The attacker is not on the path of a particular payment but it can partially observe the changes in the network. \yuvallak{3} The attacker colludes with other nodes, for example, to make packet timing analysis with sophisticated methods.

Based on these assumptions, we consider the following potential attacks for compromising privacy in PCNs: 

\begin{itemize}
    \item \textbf{Attacks on Sender/Recipient Anonymity}: Sender/Recipient anonymity requires that the identity of the sender/recipient ($u_s$/$u_r$) should not be known to the others during a payment. This is to protect the privacy of the sender/recipient so that nobody can track their shopping habits.
    %For an extreme case in a PCN even $u_r$ may not know the identity of the sender (KA: WhHAT DO YOU MEAN BY EXTREME CASE? IS IT THE CASE ALL THE TIME OR SOMETIMES? IF ALL THE TIME, THEN WE CAN REMOVE THIS SENTENCE).
    There may be cases where an adversary may successfully guess the  identity of the sender/recipient as follows: For case \yuvallak{1}, the sender can have a single connection to the network and next node is the attacker, hence, the attacker is sure that $u_s$ is sender. For case \yuvallak{2} the attacker may guess the sender/recipient by probing the changes in the channel balances. For case \yuvallak{3} the attacker will learn the sender/recipient if it can carry out a payment timing analysis within the partial network formed by the colluded nodes.
 
    %\item \textbf{Attacks on Recipient (Payee) Anonymity}: The attacks on sender anonymity also hold exactly for recipient anonymity. Therefore, we will not repeat the same explanations here. 
 %In our hypothetical setup we assume the recipient is a retail shop. Hence, the revelation of the identity of $u_r$ will partially yield the business capacity of her in a colluded attack. Since a merchant is managing a business the real identity of the merchant is known to everybody (e.g. Walmart has a known pseudonym). Hence in order to protect the trade secrets of merchant recipient privacy should be satisfied.

   \item \textbf{Attack on Channel Balance Privacy.} To keep the investment power of a user/business private, the channel capacities should be kept private in PCNs. The investment amount in a channel would give hints about financial situation of a user or its shopping preferences. Moreover, if the capacity changes in the channels are known, tracing them causes indirect privacy leakages about the senders/recipients. For instance, an attacker can initiate fake transaction requests. After gathering responses from intermediary nodes, it can learn about the channel capacities.
   %This attack is not necessarily about depleting the channel capacities but guessing the channel capacity of a node. Continuously learning the channel capacities will eventually yield more complicated privacy attacks as discussed in the attack on sender privacy.

   \item \textbf{Relationship Anonymity.} In some cases identities of $u_s$ or $u_r$ may be known. This is a very valid case for retailers because they have to advertise their identities to receive payments. However, if an attacker can relate the payer to the payee, not only the spending habits of the sender but also the the business model of the recipient will be learned. In such cases, the privacy of the trade can be preserved by hiding the relationship between the sender and recipient. Specifically, who-pays-to-whom information should be kept private. %, keeping the sender identity unidentifiable helps to protect privacy.
   %Some of the PCNs utilize onion-routing to forward the transactions to the destination node. Onion-routing is a source-routing protocol where the source of a message encapsulates the data with the keys of the intermediary nodes like a stacking doll. An intermediary node can remove only one layer from the incoming message to see the next node to which the data is to be forwarded. Hence, in a distributed network, an intermediary node will not confidently be aware of who talks to whom.
   %A network protocol may advertise the sender or the recipient during a payment. That is acceptable, required that the use case is well defined.
   %Hence, to satisfy the relationship anonymity either the recipient or the sender anonymity should be satisfied by the PCN. 
 
   \item \textbf{Business Volume Privacy.} For a retailer, publicly disclosed revenue will yield the trade secrets of its business, which must be protected by the PCN. In that sense, privacy of every payments is important.
   %single amount of payment being led to a particular merchant should stay private so that the cumulative reception amounts of the merchant can stay private. 
   %Note that for some use cases some of this private information can be inferred or the probability of guessing a piece of information can be increased. Although quantifying is really hard, we accept that being able to hide the complete picture or being able to limit the information leakage is a success for particular metrics. For example, the topology, the participating node information, the payment flowing through a path can not be hidden.
   Such payment privacy can be attacked as follows: In a scenario where two or more nodes collude, the amount of a transaction can be known to the attacker. In another scenario, if the recipient is connected to the network via a single channel through the attacker, then it will track all of the flows towards the recipient.
 
   %\item \textbf{Identity Privacy.} In PCNs channel identities are pseudonyms. However, as the routing takes place a usual internet packet forwarding the nodes have to advertise their IP addresses.
   %The IP addresses, being public, reveal the geographic location of a node. 
   %In order to mask the identity of a user, thanks to communication being TCP/IP PCNs support VPN and Tor communication.
 
\end{itemize}

\subsection{State-of-the-art PCNs and their Privacy Evaluation}

In this section, we briefly describe current studies which either present a complete PCN or propose revisions to the current ones, then analyze their privacy capabilities based on our threat model. We provide a summary of the assessment of the current PCNs' categorizations and privacy features in Table \ref{tab:comparative_table}.

\begin{table*}
\centering
\caption{Qualitative Evaluation of Privacy Features of Existing PCNs.}
\label{tab:comparative_table}
\begin{tabular}{|Z{14em}|Z{6em}|Z{3.5em}|Z{4em}|Z{4em}|Z{3.5em}|Z{4em}|Z{3.5em}|} \hline
 ~                                                   & \rot{\pbox{5em}{Network Type}} & \rot{\pbox{5em}{Blockchain Type}} & \rot{\pbox{5em}{Sender Anonymity}} & \rot{\pbox{5em}{Recipient Anonymity}} & \rot{\pbox{5em}{Channel Balance Privacy}} & \rot{\pbox{5em}{Relationship Anonymity}}     & \rot{\pbox{5em}{Business Volume Privacy}} \\ \hline \hline
 Lightning Network (HTLC) \cite{poon2016bitcoin}     & Decentralized/ Distributed                 & Public        & \halfenes    & \halfenes     & \halfenes   & \halfenes       & \fullenes       \\ \hline
 Raiden Network \cite{Raiden}                        & Decentralized/ Distributed                 & All           & \halfenes    & \halfenes     & \halfenes   & \halfenes       & \fullenes       \\ \hline
 Spider \cite{sivaraman2018high}                     & Decentralized/ Centralized                 & All           & \halfenes    & \halfenes     & \emptyenes  & \halfenes       & \halfenes       \\ \hline
 SilentWhispers \cite{malavolta2017silentwhispers}   & Decentralized/ Centralized                 & All           & \halfenes    & \halfenes     & \halfenes   & \halfenes       & \halfenes       \\ \hline
 SpeedyMurmurs \cite{roos2017settling}               & Decentralized/ Centralized                 & Public        & \halfenes    & \halfenes     & \halfenes   & \halfenes       & \halfenes       \\ \hline
 PrivPay \cite{moreno2015privacy}                    & Decentralized/ Centralized                 & Permissioned  & \halfenes    & \halfenes     & \halfenes   & \halfenes       & \halfenes       \\ \hline
% Flash \cite{wang2019flash}                         & Decentralized/ Distributed                 & Public        & \fullenes    & \halfenes     & \emptyenes  & \fullenes       & \emptyenes      \\ \hline
% Flare \cite{prihodko2016flare}                     & Distributed                                & Public        & \fullenes    & \fullenes     & \halfenes   & \fullenes       & \fullenes       \\ \hline
% Sprites \cite{miller2019sprites}                   & Centralized                                & Public        & \fullenes    & \fullenes     & \fullenes   & \fullenes       & \fullenes       \\ \hline
 Bolt \cite{green2017bolt}                           & Centralized                                & Public        & \fullenes    & \fullenes     & \fullenes   & \fullenes       & \fullenes       \\ \hline
% Fulgor \cite{malavolta2017concurrency}              & Decentralized/ Distributed                 & Public        & \fullenes    & \fullenes     & \halfenes   & \fullenes       & \fullenes       \\ \hline
% Rayo \cite{malavolta2017concurrency}                & Decentralized/ Distributed                 & Public        & \halfenes    & \halfenes     & \halfenes   & \halfenes       & \fullenes       \\ \hline
 Erdin et al. \cite{erdin2020bitcoin}                & Distributed/ Federated                     & All           & \halfenes    & \fullenes    & \halfenes    & \halfenes       & \fullenes       \\ \hline
 Anonymous Multi-Hop Locks (AMHL) \cite{malavolta2019anonymous} & Decentralized/ Distributed      & Public        & \emptyenes   & \fullenes    & \halfenes    & \fullenes       & \fullenes       \\ \hline
\end{tabular}{} 
\\
\flushleft
\hspace{10em}\halfenes : Partially satisfies OR can not defend against all mentioned attacks. \\
\hspace{10em}\fullenes : Fully satisfies. \\
\hspace{10em}\emptyenes: Does not satisfy.
\end{table*}

%\subsubsection{PCN Descriptions}
\noindent \textbf{Lightning Network (LN)}: LN \cite{poon2016bitcoin} is the first deployed PCN which utilizes Bitcoin. It started in 2017 and by June 2020 serves with more than 12.000 nodes and 36.000 channels. Nodes in LN utilize \textit{``Hashed Time-Locked Contracts''} (HTLC) for multi-hop transfer.
%money from one user to the other.
%Publicly announced payment channels are discovered by the participants through gossip messaging which means every node knows most of the topology.
The directional capacities in the payment channels are not advertised but the total capacity in the channel is known for a sender to calculate a path. This provides a partial channel balance privacy.  %The payment path is calculated by the sender. 
The sender encrypts the path by using the public keys of the intermediary nodes by utilizing \textit{``onion-routing''} so that the intermediary nodes only know the addresses of the preceding and the following nodes. None of the intermediary nodes can guess the origin or the destination of the message by looking at the network packet.% which ensures complete anonymity.
%If any of the channel capacity on the path is not sufficient, the transaction gets canceled. 

\noindent \textbf{Raiden Network}:  Shortly after LN, Ethereum foundation announced Raiden Network \cite{Raiden}. Raiden is the equivalent of LN designed for transferring Ethereum ERC20 tokens and provides the same privacy features. Although Ethereum is the second largest cryptoccurrency, that popularity is not reflected well in the Raiden Network. As of June 2020, Raiden serves with 25 nodes and 54 channels. The advantage of Raiden over LN is, due to tokenization, users can generate their own tokens to create a more flexible trading environment.

\noindent \textbf{Spider Network}: Spider network \cite{sivaraman2018routing} is a PCN which proposes applying packet-switching based routing idea which is seen in traditional networks (e.g., TCP/IP). However, it is known that in packet-switching the source and the destination of the message should be embedded in the network packet. The payment is split into many micro-payments so that the channel depletion problem gets eliminated. The authors also aimed having better-balanced channels. In this PCN, there are \textit{spider routers} with special functionalities which communicate with each other and know the capacities of the channels in the network. The sender sends the payment to a router. When the packet arrives at a router, it is queued up until the funds on candidate paths are satisfactory to resume the transaction. The authors do not mention privacy, and plan utilizing onion-routing as a future work. The micro-payments might follow separate paths, which would help keeping business volume private if the recipients were kept private. Additionally, hijack of a router will let an attacker learn everything in the network. 

\noindent \textbf{SilentWhispers}:
%The payment network in SilentWhispers \cite{malavolta2017silentwhispers} is the \textit{credit networks}, however, the idea can be well applied to the PCNs. 
SilentWhispers \cite{malavolta2017silentwhispers} utilizes landmark routing where landmarks are at the center of the payments. In their attack model, either the attacker is not on the payment path or a landmark is HBC. Here, landmarks know the topology but they do not know all of the channel balances. When sender wants to send money to a recipient, she/he communicates with the landmarks for her/his intent. Then landmarks start communicating with the possible nodes from ``sender-to-landmark'' to the ``landmark-to-recipient'' to form a payment path. Each node in the path discloses the channel balance availability for the requested transfer amount to the landmarks.
Then landmarks decide on the feasibility of the transaction by doing multi-party computation.
%During the transfer phase, when an intermediary node realizes the transaction to the next node, it informs the landmark.
%Landmarks acknowledge the transactions and when all of the transactions are executed on the intended path, the transaction is marked successful.
In SilentWhispers, the sender and the receiver are kept private but the landmarks know the sender-recipient pair. The payment amount is also private for the nodes who do not take part in the transaction. Moreover, the balances of the channels within the network are kept private. Although centralization is possible, the approach is decentralized and landmarks are trusted parties.

\noindent \textbf{SpeedyMurmurs}:  %This paper lacks a lot of things.
SpeedyMurmurs \cite{roos2017settling} is a routing protocol, specifically an improvement for LN. In SpeedyMurmurs, there are well-known landmarks like in SilentWhispers. The difference of this approach is that the nodes on a candidate path exchange their neighbors' information anonymously. So if a node is aware of a path closer to the recipient, it forwards the payment in that direction, called ``shortcut path''. In a shortcut path, an intermediary node does not necessarily know the recipient but knows a neighbor close to the recipient. %Another point in the study is that the payment routing is divided into small chunks and forwarded to the recipient.
%They talk about value privacy, sender privacy, receiver privacy. Value privacy is similar to that of HTLC. 
SpeedyMurmurs hides the identities of the sender and the recipient by generating anonymous addresses for them. Intermediary nodes also hide the identities of their neighbors by generating anonymous addresses.
%Based on their assumption sender and receiver privacy is hard to achieve.
Although it may be complex, applying de-anonymization attacks on the network will turn it into SilentWhispers. This is because, while the algorithm is a decentralized approach, with unfair role distribution, it may turn into a centralized approach.
%They send chunks to the landmarks. The sender will behave like a landmark. it is obvious that she is the sender. The routing is not suitable for honest-but-curious attackers. Because the routing is basically packet-switch based. They say PCNs resemble F2F networks. So there is a trust going on. It is not clear how privacy is achieved. 

\noindent \textbf{PrivPay}: PrivPay \cite{moreno2015privacy} is a hardware-oriented version of SilentWhispers. The calculations in the landmark are done in tamper-proof trusted hardware. Hence, the security and privacy of the network are directly related to the soundness of the trusted hardware which may also bring centralization.
%Furthermore, the idea is presented to be applied in the credit networks but it is still applicable for PCNs.
%The trusted hardware is in the center of operations and it is managed by a central node that yields high centrality. 
In PrivPay, sender privacy is not considered. Receiver privacy and business volume privacy is achieved by misinformation. When an attacker constantly tries to query data from other nodes the framework starts to produce probabilistic results.

\noindent \textbf{Bolt:} Bolt \cite{green2017bolt} is a hub-based payment system. That is, there is only one intermediary node between sender and recipient. Bolt assumes \textit{zero-knowledge proof} based cryptocurrencies. It does not satisfy privacy in multi-hop payments, however, it satisfies very strong relationship anonymity if the intermediary node is honest. On the other hand, being dependant on a single node makes this approach a centralized one. 
%\paraphenes{Green and Miers presented BOLT, a hub-based privacy-preserving payment for PCNs [32]. BOLT requires cryptographic primitives only available in Zcash and it cannot be seamlessly deployed in Bitcoin. Moreover, this approach is limited to paths with a single intermediary and the extension to support paths of arbitrary length remains an open problem}

\noindent \textbf{Permissioned Bitcoin PCN}: In PCNs, %the secure communication protocols are undoubtedly important, however, 
if the network topology is not ideal, e.g., star topology, some of the nodes may learn about the users and payments. To this end, the authors in \cite{erdin2020bitcoin} propose a new topological design for a permissioned PCN % on how the PCN topology should be established 
such that the channels' depletion can be prevented. They come up with a real use case where %shoppers pay to the merchants. They propose that the 
a consortium of merchants create a full P2P topology and the customers connect to this PCN through merchants which undertakes the financial load of the network to earn money. The privacy of the users in the PCN is satisfied by LN-like mechanisms. The authors also investigate how initial channel balances change while the sender/receiver privacy and the relationship anonymity can be satisfied %more strongly 
by enforcing at least 3-hops in a multi-hop payment. 

\noindent \textbf{Anonymous Multi-Hop Locks (AMHL)}: In AMHL proposal \cite{malavolta2019anonymous}, the authors offer a new HTLC mechanism for PCNs. On a payment path, the sender agrees to pay some service fee to each of the intermediaries for their service. However, if two of these intermediaries maliciously collude they can eliminate honest users in the path and consequently steal their fees. In order to solve this, they introduce another communication phase in which the sender distributes a one-time-key to the intermediary nodes. Although the HTLC mechanism is improved for the security of the users the sender's privacy is not protected; each of the intermediaries learns the sender. However, relationship anonymity can still be satisfied.

%\paraphenes{THIS IS FROM THEIR PAPER Relationship Anonymity. Relationship anonymity [12] requires that each intermediate node does not learn any information about the set of users in an AMHL beyond its direct neighbors. This property is satisfied by Fasthe lock identifiers are sampled at random and during the locking phase, a user only learns the identifiers of its left and right lock as well as its left and right neighbor. We discuss this further in Appendix}

%\textbf{Coinexpress}: \cite{yu2018coinexpress} privacy is lacking perhaps no need to mention.

% \notenes{SM: It seems that privacy and efficiency are trade-offs when designing a PCN. (This can be commented somewhere if needed)}

%\newpage%===========================================
%===================================================
%\section{Privacy in Payment Channel Networks}
%\section{Investigation of Privacy in PCNs}

%     /$$$$$$                        /$$     /$$                    
%    /$$__  $$                      | $$    |__/                    
%   | $$  \__/  /$$$$$$   /$$$$$$$ /$$$$$$   /$$  /$$$$$$  /$$$$$$$ 
%   |  $$$$$$  /$$__  $$ /$$_____/|_  $$_/  | $$ /$$__  $$| $$__  $$
%    \____  $$| $$$$$$$$| $$        | $$    | $$| $$  \ $$| $$  \ $$
%    /$$  \ $$| $$_____/| $$        | $$ /$$| $$| $$  | $$| $$  | $$
%   |  $$$$$$/|  $$$$$$$|  $$$$$$$  |  $$$$/| $$|  $$$$$$/| $$  | $$
%    \______/  \_______/ \_______/   \___/  |__/ \______/ |__/  |__/
%
%\newpage%==========================================
%==================================================
\section{Future Research Issues in PCNs}
\label{sec:FutureResearch}

%Cryptocurrencies, especially the Bitcoin, redefined how money can be kept and transferred from one to another. However, public blockchain-based cryptocurrencies suffer from high transaction wait times and absurd transaction fees which are independent of the amount being sent. These drawbacks hinder the wide usage of cryptocurrencies. Among many, payment channel network idea is the most accepted solution for cryptocurrencies to be used for micro-payments. This idea fixed the mentioned shortcomings: payment approval measured in seconds, fees proportional to the payment amount. There are many studies on how payment channels and payment channel networks should be designed to make the transfers secure and efficient. However, these studies do not mention the possible privacy leakages of these methods in case of a wide adaptation of proposed ideas. In this paper, we propose a categorization of the networks formed by the blockchain-backed cryptocurrencies. After demonstrating possible attacks on user privacy in the payment channel networks, we classify the state-of-the-art payment channel networks based on what they propose in terms of privacy. Based on the discussions on the strengths and weaknesses of the approaches, we offer possible directions for research for the future of privacy based payment channel networks. 

%PAYMENT NETWORK NASIL OLMALI?
Privacy in PCNs is an understudied topic and there are many open issues that need to be addressed as a future research. In this section, we summarize these issues:

\noindent \textbf{Abuse of the PCN protocols.} As most of the PCNs rely on public cryptocurrencies, whose protocol implementations are public. This freedom can be abused such that by changing some parameters and algorithms in the design, an attacker can behave differently than what is expected. % e.g., sybil attack, sink hole attack etc. 
This will bring privacy leakages and censorship in the network. A topological reordering of the network will help solve this problem. If a sender gets suspicious about an intermediary node, it can look for alternatives instead of using that node.

%Although many studies are offering both theoretical and proof-of-concept solutions for the PCNs, 

%\textbf{PCN topologies.} The most widely accepted and readily available solution, Lightning Network, has a user base of around 10 \textcolor{red}{AK: 12k} thousand nodes as of today. Furthermore, if the channels are observed it creates an impression that most of the nodes are experimental to discover the capabilities of LN. Even the trust on the protocol becomes perfect, assuming that ordinary users will put hundreds of dollars in their channels as collateral does not make perfect sense. This reality reminds us that PCNs are inclined to slide towards centrally managed networks. In that case, topology formation comes in the scene. Right now, the \texttt{autopilot} feature of \texttt{lnd} (an LN client) highlights scale-free Barabasi-Albert network formation method. However, this method does not take the financial strength of the attendees but only their existence.

\noindent \textbf{Discovery of Colluding Nodes.} %Another threat for a possible leak of user privacy is the colluding nodes. 
When the nodes collude in a PCN, they can extract more information about the users. To prevent this, the protocols should be enriched to discover the colluding nodes or by adding redundancy to the protocols, colluding nodes can be confused.

\noindent \textbf{Policy Development.} The cryptocurrency and PCN idea is still in the early phases of their lives. Hence, policy and regulation for not only the security of the participants but also for the privacy of them is highly needed in this domain. This will also create a quantitative metric for the researchers to measure the success of their proposals.

\noindent \textbf{Impact of Scalability on Privacy.} One of the aims for introducing PCNs was making the cryptocurrencies more scalable. For example, LN advises running the Barabasi-Albert scale-free network model while establishing new connections \cite{martinazzi2020evolving}. Thus, the final state of the network can impose centralization which will have adverse effects on the privacy of the nodes in the network.
%Therefore, there is a need to study topology expansions in a privacy-preserving  manner. 

\noindent \textbf{Integration of IoTs with PCNs.} Use of IoT devices for payments are inevitable. Aside from the fact that most IoT devices are not powerful to run a full node, security and privacy of the payments and the device identities within the IoT ecosystem needs to be studied. These devices are anticipated to be able to participate in the network through gateways. Revelation of device ownership will reveal the real identity of the users to the public which is a big threat on privacy.

\noindent \textbf{Privacy in Permissioned PCNs.} While establishing a network of merchants in permissioned PCNs, the merchants should at least disclose their expected trade volume in order to establish a dependable network. This will, however, yield trade secrets of the merchants. To prevent this, zero-knowledge proof based multi-party communication can be explored.
%\noindent {PRIVACY IN PERMISSIONED PCNs. IF YOU VOTE ON TRANSACTIONS, THEN YOU WILL HAVE ACCESS TO THEIR INFO?.}\\

%Regulations and Liability, auditing??

%\textbf{Atomic Multi-path Payments.}

%We can get some information from Ahmet. To-do work.
%\textcolor{red}{AK: LN recently introduced multi-part payment sending, maybe can be mentioned somewhere in the paper since this benefits payment privacy. https://github.com/lightningnetwork/lnd/pull/3967, https://lightning.engineering/posts/2020-05-07-mpp/}

%%%%%%%%%%%%%%%%%%%%%%%%%%%%%%%%%%%%%%
%%%%%%%%%%%%%%%%%%%%%%%%%%%%%%%%%%%%%%
%%&$&$&$&$&$&$&$&$&$&$&$&$&$&$&$&$&$&$
%     /$$$$$$                        /$$     /$$                    
%    /$$__  $$                      | $$    |__/                    
%   | $$  \__/  /$$$$$$   /$$$$$$$ /$$$$$$   /$$  /$$$$$$  /$$$$$$$ 
%   |  $$$$$$  /$$__  $$ /$$_____/|_  $$_/  | $$ /$$__  $$| $$__  $$
%    \____  $$| $$$$$$$$| $$        | $$    | $$| $$  \ $$| $$  \ $$
%    /$$  \ $$| $$_____/| $$        | $$ /$$| $$| $$  | $$| $$  | $$
%   |  $$$$$$/|  $$$$$$$|  $$$$$$$  |  $$$$/| $$|  $$$$$$/| $$  | $$
%    \______/  \_______/ \_______/   \___/  |__/ \______/ |__/  |__/
%
\section{Conclusion}
\label{sec:conclusion}
PCN is a promising solution to make the cryptocurrency-based payments scalable. This idea aimed fixing two major shortcomings of the cryptocurrencies: long confirmation times and high transaction fees.
%by decreasing payment approval times and decreasing the transfer fees  and making them proportional to the payment amount.
There are many studies on the design of payment channels and PCNs to make the transfers secure and efficient. However, these studies do not mention the possible privacy leakages of these methods in case of a wide adaptation of proposed ideas. In this paper, we first made the categorization of PCNs based on the type of blockchain being used and the topological behavior of the network. After clearly defining possible privacy leakages in a PCN, we compared and contrasted the state-of-the-art PCN approaches from the privacy point of view.

%\newpage%===========================================
\bibliographystyle{unsrt}

\end{document}